\documentclass[reprint,superscriptaddress,nobibnotes,amsmath,amssymb,aps,prx,hidelinks]{revtex4-2}

\usepackage{showyourwork}
\usepackage{graphicx}
\usepackage{dcolumn}
\usepackage{bm}
\usepackage{multirow}
\usepackage[separate-uncertainty=true]{siunitx}
\usepackage{amsmath}
\usepackage[shortlabels]{enumitem}
\usepackage[page]{appendix}
\usepackage{xfrac}
\usepackage{printlen}
\usepackage[version=4]{mhchem}
\usepackage{blindtext}
\usepackage[noabbrev,nameinlink,capitalize]{cleveref}
\crefname{equation}{Eqn.}{Eqns.}
\crefformat{equation}{Eqn.~#2#1#3}
\crefrangeformat{equation}{Eqns.~#3#1#4 to~#5#2#6}
\crefmultiformat{equation}{Eqns.~#2#1#3}{ and~#2#1#3}{, #2#1#3}{ and~#2#1#3}
\crefname{figure}{Fig.}{Figs.}

\newcommand{\papertitle}{Accurate Estimation of Diffusion Coefficients and their Uncertainties from Computer Simulation}

\newcommand{\oMSD}{\ensuremath{\bm{x}}}
\newcommand{\oMSDs}[1]{\ensuremath{x}(#1)}
\newcommand{\oMSDi}{\ensuremath{x_i}}
\newcommand{\oMSDj}{\ensuremath{x_j}}
\newcommand{\oMSDn}{\ensuremath{x_n}}
\newcommand{\oMSDm}{\ensuremath{x_{n + m}}}
\newcommand{\moMSDn}{\ensuremath{\left<\oMSDn\right>}}
\newcommand{\moMSDm}{\ensuremath{\left<\oMSDm\right>}}
\newcommand{\modelmatrix}{\mathbf{A}}
\newcommand{\model}{\bm{m}}
\newcommand{\prob}[1]{\ensuremath{p(#1)}}

\newcommand{\nind}[1]{\ensuremath{N^\prime_{#1}}}
\newcommand{\code}[1]{#1}

\newcommand{\MSD}[1]{\big\langle\Delta\mathbf{r}{(#1)}^2\big\rangle}
\newcommand{\Dest}{\ensuremath{\widehat{D}^*}}
\newcommand{\D}{\ensuremath{D^*}}
\newcommand{\var}[1]{\ensuremath{\sigma^2[#1]}}
\newcommand{\varest}[1]{\ensuremath{\widehat{\sigma}^2[#1]}}
\newcommand*\xbar[1]{
  \hbox{
    \vbox{
      \hrule height 0.5pt 
      \kern0.5ex
      \hbox{
        \kern-0.1em
        \ensuremath{#1}
        \kern-0.1em
      }
    }
  }
}

\graphicspath{{figures/}}

\makeatletter
\def\maketitle{
\@author@finish
\title@column\titleblock@produce
\suppressfloats[t]}
\makeatother

\begin{document}

\let\oldaddcontentsline\addcontentsline
\renewcommand{\addcontentsline}[3]{}

\title{\papertitle}

\author{Andrew R. McCluskey}
\email{andrew.mccluskey@bristol.ac.uk}
  \affiliation{Centre for Computational Chemistry, School of Chemistry, University of Bristol, Cantock's Close, Bristol, BS8 1TS, UK.}
  \affiliation{European Spallation Source ERIC, Data Management and Software Centre, Asmussens Allé 305, DK-2800 Kongens Lyngby, DK.}
  \affiliation{Diamond Light Source, Harwell Campus, Didcot, OX11 0DE, UK.}
\author{Samuel W. Coles}
  \affiliation{Department of Chemistry, University of Bath, Claverton Down, Bath, BA2 7AY, UK}
  \affiliation{The Faraday Institution, Quad One, Harwell Science and Innovation Campus, Didcot, OX11 0RA, UK}
\author{Benjamin J. Morgan}
  \email{b.j.morgan@bath.ac.uk}
  \affiliation{Department of Chemistry, University of Bath, Claverton Down, Bath, BA2 7AY, UK}
  \affiliation{The Faraday Institution, Quad One, Harwell Science and Innovation Campus, Didcot, OX11 0RA, UK}

\begin{abstract}
    Self-diffusion coefficients, $\D$, are routinely estimated from molecular dynamics simulations by fitting a linear model to the observed mean-squared displacements (MSDs) of mobile species.
    MSDs derived from simulation exhibit statistical noise that causes uncertainty in the resulting estimate of $\D$.
    An optimal scheme for estimating $\D$ minimises this uncertainty, i.e., it will have high statistical efficiency, and also gives an accurate estimate of the uncertainty itself.
    We present a scheme for estimating $\D$ from a single simulation trajectory with high statistical efficiency and accurately estimating the uncertainty in the predicted value.
    The statistical distribution of MSDs observable from a given simulation is modelled as a multivariate normal distribution using an analytical covariance matrix for an equivalent system of freely diffusing particles, which we parameterise from the available simulation data. 
    We use Bayesian regression to sample the distribution of linear models that are compatible with this multivariate normal distribution, to obtain a statistically efficient estimate of $D^*$ and an accurate estimate of the associated statistical uncertainty.
\end{abstract}

\maketitle 

\section{Introduction}

Mass transport is a fundamental physical process that is central to our understanding of fluids~\cite{sendner_interfacial_2009,ShimizuEtAl_PhysChemChemPhys2015,ghoufi_ultrafast_2016} and plays a critical role in biochemical systems~\cite{maccmmon_dynamics_1977,RobertsonEtAl_ProcNatlAcadSci2006} and solid-state devices, such as batteries, fuel cells, and chemical sensors~\cite{eames_ionic_2015,morgan_understanding_2011,walsh_taking_2018}.
Molecular dynamics simulations are widely used to study microscopic transport processes, as they give direct insight into atomic-scale transport mechanisms and can be used to calculate macroscopic transport coefficients~\cite{morgan_relationships_2014,morgan_mechanistic_2021,poletayev_defect_2022,klepis_long_2009,wang_application_2011,ZelovichEtAl_ChemMater2019}.
These transport coefficients are formally defined in terms of ensemble averages.
Dynamical simulations, however, sample the full ensemble space stochastically, and parameters derived from simulation data, therefore, are estimates of the true parameter of interest.
The statistical uncertainty associated with such estimates depends on the details of the simulation---e.g., size and timescale---and on the choice of estimation method.
An optimal estimation method will minimise the uncertainty in the computed quantity, i.e., it will have high statistical efficiency, and will also allow this uncertainty to be accurately estimated.

One commonly used parameter for quantifying atomic-scale mass transport is the self-diffusion coefficient, $\D$, which describes diffusion in the absence of a chemical potential gradient.
$\D$ is related to the ensemble-average mean squared displacement (MSD), $\MSD{t}$, via the Einstein relation~\cite{einstein_uber_1905,helfand_transport_1960},
\begin{equation}
    \D = \lim_{t\to\infty}\frac{\MSD{t}}{6t},
    \label{equ:einstein}
\end{equation}
where $\Delta\mathbf{r}{(t)}$ is the displacement of a diffusing particle in the time interval $t$.
Because numerical simulations are finite in time and space, MSDs obtained from simulation data always differ from the true ensemble average MSD.
One can, however, compute an estimate of the self-diffusion coefficient, $\Dest$, by fitting a linear model to the observed MSD and using the gradient of this fitted model in place of $\MSD{t} / t$ in \cref{equ:einstein}~\footnote{While estimating $\D$ by fitting a linear model to MSD data is, perhaps, the most commonly used approach in the literature, alternative methods that avoid linear fitting of MSD data also exist; see, for example, Refs.~{\protect\cite{VestergaardEtAl_PhysRevE2014,KrapfEtAl_NewJournalPhysics2018,bullerjohn_optimal_2020}}.}.

The simplest approach to fitting a linear model to MSD data from simulation is ordinary least squares regression (OLS).
OLS gives analytical expressions for the ``best fit'' regression coefficients (the slope and intercept) and their respective uncertainties, making it easy to implement and quick to perform.
OLS, however, is appropriate only for data that are both statistically independent and identically distributed.
Neither of these conditions holds for MSD data obtained from simulation, which instead are serially correlated and usually have unequal variances.
As a consequence, OLS is statistically inefficient, giving a relatively large statistical uncertainty in $\Dest$.
Furthermore, the textbook OLS expression for the uncertainty in $\Dest$ significantly underestimates the true uncertainty in this estimate~\cite{UslerEtAl_JComputChem2023}.
This underestimated uncertainty may give overconfidence in the accuracy of values of $\D$ estimated using OLS, and propagating these uncertainties in any downstream analyses may result in faulty inferences.
While the uncertainty associated with OLS estimates of $\D$ can, in principle, be accurately estimated by directly sampling over multiple repeated simulations, this approach can greatly increase the total computational cost and is often impractical.

Here, we describe an approximate Bayesian regression method for estimating $\D$ with near-maximal statistical efficiency while accurately estimating the corresponding statistical uncertainty using data from a single simulation.
We model the statistical population of simulation MSDs as a multivariate normal distribution, using an analytical covariance matrix derived for an equivalent system of freely diffusing particles, with this covariance matrix parameterised from the observed simulation data.
We then use Markov-chain Monte Carlo to sample the posterior distribution of linear models compatible with this multivariate normal model.
The resulting posterior distribution provides an efficient estimate for $\D$ and allows the associated statistical uncertainty in $\Dest$ to be accurately quantified.
This method is implemented in the open-source Python package \textsc{kinisi}~\cite{mccluskey_kinisi_2022}.

\section{Background}

For a simulation of equivalent particles, the observed mean squared displacement as a function of time, $\oMSDs{t}$, can be computed as an average over equivalent particles and time origins:
\begin{equation}
  \oMSDs{t} = \frac{1}{N(t)}\sum^{N(t)}_{j=1}{\left[\Delta\mathbf{r}_j(t)\right]}^2,
  \label{equ:observed_msd}
\end{equation}
where $N(t)$ is the total number of observed squared-displacements at time $t$.
The resulting observed MSD is a vector, $\oMSD$, with individual elements $\oMSDi$.
Each element of this vector differs from the true ensemble-average MSD for that time by some unknown amount.
Fitting a linear model to $\oMSD$ gives an estimated self-diffusion coefficient, $\Dest$, which again differs from the true self-diffusion coefficient, $\D$, by some unknown amount.

Performing repeated simulations starting from different random seeds or with different histories will produce a set of replica trajectories, where each trajectory gives a different, statistically equivalent, observed MSD.
The set of all possible replica trajectories defines a population of hypothetical observed MSDs, and the MSD obtained from any one trajectory can be considered a random sample, $\bm{X}$, drawn from the multivariate probability distribution that describes this population, i.e, $\bm{X} \sim \prob{\oMSD}$.
Each potential MSD sample could, in principle, be fitted to a linear model to obtain a corresponding estimate for the self-diffusion coefficient; $\bm{X} \mapsto \Dest$.
The population of all such estimates, therefore, defines a probability distribution $\prob{\Dest}$.
The estimated diffusion coefficient obtained from a single simulation corresponds to a random sample drawn from this distribution, while the uncertainty in $\Dest$ is described by the shape of the full distribution $\prob{\Dest}$.

The statistical properties of $p(\Dest)$ depend on both the input MSD data and the choice of regression scheme used to obtain a ``best fit'' linear model.
An optimal estimation scheme for $\D$ should be unbiased, i.e., the expected value, $\mathbb{E}(\Dest)$, should equal the true self-diffusion coefficient $\D$, and should be maximally statistically efficient, i.e., the spread of $p(\Dest)$ around $\D$ should be minimised.
An estimation scheme should also provide an accurate estimate of the uncertainty in $\Dest$, to allow this estimated parameter to be used in subsequent inferential analysis.

\begin{figure}
    \centering
    \resizebox{\columnwidth}{!}{\includegraphics*{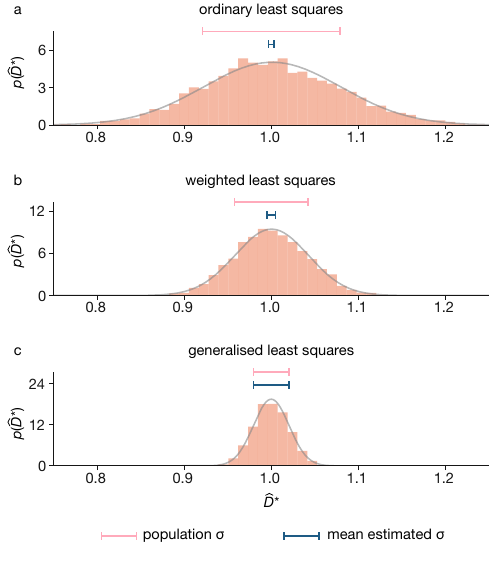}}
    \caption{
        Example distributions of estimated self-diffusion coefficients, $\Dest$, calculated using (a) ordinary least squares (OLS), (b) weighted least squares (WLS), and (c) generalised least squares (GLS),
        from MSD data from \num{4096} individual simulations of \num{128} particles undergoing a \SI{128}{step} 3D lattice random walk, with a step size chosen so that the true diffusion coefficient $\D = 1$.
        In each panel, the grey curve shows the best-fit normal distribution for the simulation data, the upper horizontal bar shows the standard deviation of this distribution, and the lower horizontal bar shows the average estimated standard deviation given by the analytical expression for $\sigma[\prob{\Dest}]$ for each regression method.}
    \label{fig:glswlsols}
    \script{glswlsols.py}
\end{figure}

For data that are both statistically independent and identically normally distributed, ordinary least squares regression (OLS) is unbiased and statistically efficient, and gives accurate estimates of the uncertainties in the resulting regression coefficients.
MSD data obtained from simulation, however, are neither statistically independent nor identically distributed.
The variances, $\var{\oMSDi}$, are correlated, since the displacement of each particle at time $t+\Delta t$ is necessarily similar to its displacement at time $t$, and hence, $\oMSDs{t}$ is similar to $\oMSDs{t+\Delta t}$.
These variances are also typically unequal---the data are heteroscedastic \cite{smith_random_1996,he_statistical_2018,UslerEtAl_JComputChem2023}. 
Because the key assumptions of the OLS method are not valid for MSD data, OLS gives statistically inefficient estimates of $\D$, while the estimated regression uncertainties obtained from the standard OLS statistical formulae significantly underestimate the true uncertainty in $\prob{\Dest_\mathrm{OLS}}$ (\cref{fig:glswlsols}a).

Some improvement can be made by using weighted least squares (WLS) (\cref{fig:glswlsols}b), where the residual for each observed MSD value is weighted by the reciprocal of its variance, $1/(\var{\oMSDi})$.
Like OLS, WLS is an unbiased estimator, and for heteroscedastic data it has higher statistical efficiency than OLS.
WLS still neglects correlations in $\oMSD$, however, and is therefore statistically inefficient, and the WLS estimated uncertainties for the regression coefficients still underestimate the true uncertainty in $\prob{\Dest_\mathrm{WLS}}$.

To optimally estimate the true ensemble-average MSD, and hence $\D$, from simulation data, it is necessary to account for both the changing variance and correlation structure of $\oMSD$.
Within the framework of linear regression, this can be achieved using generalised least squares (GLS).
GLS gives estimated regression coefficients, $\widehat{\beta}$, via
\begin{equation}
    \widehat{\beta} = \left(\modelmatrix^{\top}\mathbf{\Sigma}^{-1}\modelmatrix\right)^{-1}\modelmatrix^{\top}\mathbf{\Sigma}^{-1}\oMSD,
    \label{equ:gls}
\end{equation} 
where $\modelmatrix$ is the model matrix $\begin{bmatrix}\mathbf{1} & \bm{t}\end{bmatrix}$, with $\bm{t}$ the vector of observed times, and $\mathbf{\Sigma}$ is the covariance matrix for the observed MSD values.
For correlated heteroscedastic data, such as MSD data, GLS offers the theoretical maximum statistical efficiency---it achieves the Cram\'er--Rao bound~\cite{cramer_mathematical_1946,rao_information_1945,rao_selected_1994,darmois_sur_1945,aitken_on_1942}---and provides accurate analytical estimates of the uncertainty in the predicted regression coefficients (\cref{fig:glswlsols}c).

An alternative method for estimating the ensemble-average MSD, and thus $\Dest$, from simulation data is Bayesian regression.
Like GLS, Bayesian regression can take into account both the changing variance and the correlation structure inherent in the data.
Rather than providing a singular ``best-fit'' estimate like GLS, Bayesian regression produces a posterior joint probability distribution for the regression coefficients.
The mean of this distribution serves as a point estimate of the coefficients and, in the absence of additional prior information, is equal to the estimate obtained from GLS, while the spread of the distribution quantifies the uncertainty in these estimates.
For data that are both heteroscedastic and correlated, such as MSD data from simulations, Bayesian regression, like GLS, is formally fully statistically efficient.

The estimation of $\D$ from some observed MSD data, $\oMSD$, using Bayesian regression, proceeds by computing the posterior probability distribution $\prob{\model|\oMSD}$ for a linear model $\model = 6\D \bm{t} + c$, where $\D$ and $c$ are parameters to be estimated.
This posterior distribution is described by Bayes' theorem,
\begin{equation}
    \prob{\model|\oMSD} = \frac{\prob{{\oMSD|\model}}\prob{\model}}{\prob{\oMSD}},
    \label{equ:bayes}
\end{equation}
where $\prob{\oMSD|\model}$ is the probability of observing data $\oMSD$ given model $\model$, often described as the ``likelihood'', and $\prob{\oMSD}$ is the marginal probability of the observed data $\oMSD$.
Integrating over $\prob{\model|\oMSD}$ with respect to $c$ yields the marginal posterior distribution $\prob{\D|\oMSD}$, from which the best point-estimate $\Dest$ and distribution variance $\varest{\Dest}$ can be computed.

Given a sufficiently large number of observed squared displacements at each time $t$, the central limit theorem applies, and $\oMSD$ can be considered a sample from a multivariate normal distribution with log-likelihood
\begin{equation}
    \begin{aligned}{}
      \ln \prob{\oMSD|\model} = -\frac{1}{2}\big[ & \ln(\left|\mathbf{\Sigma}\right|) + {(\oMSD - \model)}^{\!\top}\mathbf{\Sigma}^{-1}(\oMSD - \model) \\ 
      & + k \ln(2\pi)\big],
    \end{aligned}
    \label{equ:loglike}
\end{equation}
where $\mathbf{\Sigma}$ is the observed MSD covariance matrix and $k$ is the length of the vector $\oMSD$, i.e., the number of time intervals for which we have observed MSD data.
Providing that this likelihood function can be calculated, one can compute the posterior distribution $\prob{\model | \oMSD}$ via \cref{equ:bayes}, to obtain an optimally efficient point-estimate for $\D$ and a complete description of the associated uncertainty in $\Dest$.

\section{Approximating $\bm{\Sigma}$ from simulation data}

The practical application of Bayesian regression or GLS requires the covariance matrix for the observed MSD, $\mathbf{\Sigma}$, which is generally unknown.
To proceed, we approximate $\mathbf{\Sigma}$ with a model covariance matrix, $\mathbf{\Sigma^\prime}$, with a known analytical form, that we parameterise from the available simulation data.
Providing the correlation structure of $\mathbf{\Sigma^\prime}$ is similar to that of $\mathbf{\Sigma}$, this model correlation matrix can be used in approximate Bayesian or GLS schemes to estimate the ensemble-average MSD, and hence $\D$, with high efficiency and accurate estimated uncertainties.

We model the covariance matrix for the observed MSD from a given simulation using the covariance matrix for the MSD of an equivalent system of freely diffusing particles, $\mathbf{\Sigma^\prime}$.
We note that estimating \Dest\ by fitting a linear model to observed MSD data implicity assumes that these data sample the long-time limit where the Einstein relation (\cref{equ:einstein}) is valid.
In this long-time diffusive regime, all systems of mobile particles are statistically equivalent under rescaling by \D, and hence have the same MSD covariance structure $\mathbf{\Sigma}$. 

For observed MSDs computed by averaging over non-overlapping time windows, the covariance matrix $\mathbf{\Sigma^\prime}$, in the long-time limit, has elements (see SI)
\begin{equation}
  \Sigma^\prime\left[\oMSDi, \oMSDj\right]= \Sigma^\prime\left[\oMSDj, \oMSDi\right] =
  \var{\oMSDi} \frac{\nind{i}}{\nind{j}},\hspace{1em} \forall\,i \leq j,
  \label{equ:cvv}
\end{equation} 
where $\var{\oMSDi}$ are the time-dependent variances of the observed MSD, and $\nind{i}$ is the total number of non-overlapping observed squared-displacements for time-interval $i$.
We estimate the variances $\var{\oMSDi}$ using the standard result that the variance of the mean of a sample scales inversely with the number of independent constituent observations.
Specifically, we compute an estimated variance $\varest{\oMSDi}$ by rescaling the observed variance of the squared displacement for time interval $i$ by the number of numerically-independent contributing sub-trajectories, $\nind{i}$, which is given by the number of mutually non-overlapping time windows of length $i$ multiplied by the number of mobile particles, summed over all simulations used to compute the MSD;
\begin{equation}
  \varest{\oMSDi} = \frac{1}{\nind{i}}\var{\Delta r_i^2}.
  \label{equ:varestMSD}
\end{equation}
By rescaling by the number of numerically-independent contributing sub-trajectories, rather than by the total number of observed squared displacements for time window $i$, we account for correlations between the squared displacements of each particle computed from overlapping time windows (further details are provided in the SI).

\begin{figure}
    \centering
    \resizebox{\columnwidth}{!}{\includegraphics*{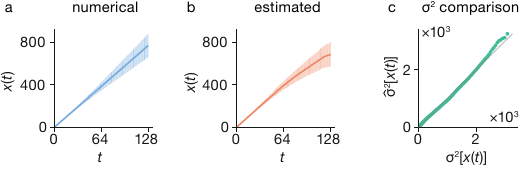}}
    \caption{
        Comparison of the numerical variance in observed MSD from multiple replica simulations and the estimated variance in observed MSD given by rescaling the variance in observed squared displacements (\cref{equ:varestMSD}).
        Panel (a) shows the mean observed MSD from \num{4096} simulations of \num{128} particles undergoing a 3D lattice random walk of \num{128} steps per particle, with error bars of $\pm2\sigma[\oMSDi]$.
        Panel (b) shows the MSD from just one simulation, with error bars of $\pm2\widehat{\sigma}[\oMSDi]$, obtained via \cref{equ:varestMSD}.
        Panel (c) plots the numerical variance against the estimated variance from a single simulation as a function of timestep $i$.
    }
    \label{fig:msd}
    \script{msd.py}
\end{figure}

The estimated variance $\varest{\oMSD}$ can be calculated from a single simulation trajectory, and provides an accurate estimate of the true variance $\var{\oMSD}$~\footnote{A common approach to estimating the variance of the mean of time-correlated data with unknown correlation time is the renormalization group blocking method of Flyvberg and Peterson~\cite{FlyvbjergAndPetersen_JChemPhys1989,Frenkel2023-ah}. In the SI, we compare this blocking method to our direct rescaling method (\cref{equ:varestMSD}) for estimating $\var{\oMSDi}$ from a set of random-walk trajectories. For our example data, direct rescaling performs better (gives more accurate estimates of $\var{\oMSDi}$).}.
To demonstrate this, we performed \num{4096} independent simulations of \num{128} particles undergoing a three-dimensional cubic-lattice random walk of \num{128} steps per particle.
Using data from all \num{4096} simulations, we first compute the true simulation MSD and its variance (\cref{fig:msd}a).
We also compute the MSD and estimated variance using data from a single simulation trajectory (\cref{fig:msd}b), using the scheme described above.
A quantitative comparison between the true MSD variance and the single-trajectory estimated MSD variance is made in \cref{fig:msd}c: the close numerical agreement confirms that \cref{equ:varestMSD} can be used to estimate $\var{\oMSD}$, which can then be used to parameterise the model covariance matrix $\mathbf{\Sigma^\prime}$ via \cref{equ:cvv}.

The practical implementation of both GLS and Bayesian regression requires that the covariance matrix $\mathbf{\Sigma^\prime}$ is invertible (positive definite); see \cref{equ:gls,equ:loglike}.
The estimated MSD variances derived from simulation data via \cref{equ:varestMSD}, however, are statistically noisy and using these to directly parameterise $\mathbf{\Sigma^\prime}$ can yield matrices with high condition numbers, resulting in numerical instabilities when these are used in \cref{equ:loglike}, or matrices that are singular and non-invertible.
To make our scheme numerically tractable, we recondition the estimated covariance matrix obtained from \cref{equ:cvv} using the minimum eigenvalue method~\cite{TabeartEtAl_TellusDynMeteorolOceanogr2020}.
This approach ensures that the condition number for the resulting covariance matrix is equal to a user defined parameter that can be set to produce an invertible matrix that allows numerically stable GLS or Bayesian regression.

\begin{figure}
    \centering
    \resizebox{\columnwidth}{!}{\includegraphics*{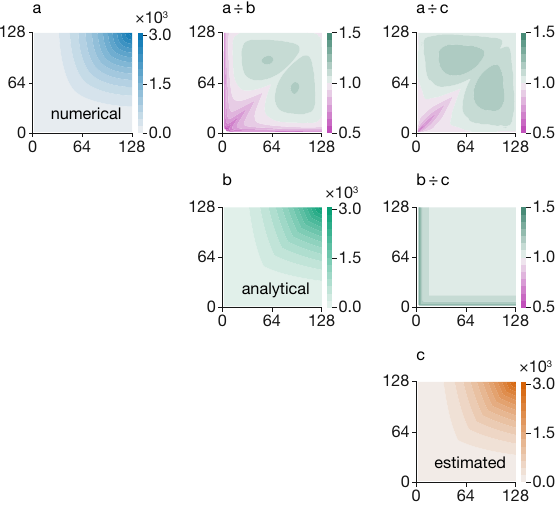}}
    \caption{
        (a) The numerical MSD covariance matrix $\bm{\Sigma}$ calculated using MSD data from \num{4096} simulations of \num{128} particles undergoing a 3D lattice random walk of \num{128} steps per particle.
        (b) The analytical MSD covariance matrix $\bm{\Sigma^\prime}$ (\cref{equ:cvv}), parametrised using analytical long-time limit random-walk variances $\var{\oMSDi}$.
        (c) The MSD covariance matrix obtained applying the numerical scheme described in the main text to each individual random walk simulation, averaged over all \num{4096} such simulations.
        Colour bars in (a--c) show the covariance, $\Sigma\left[\oMSDi, \oMSDj\right]$.
        The off-diagonal panels show difference plots, computed as per-element ratios between pairs of covariance matrices (a--c).
    }
    \label{fig:covariances}
    \script{covariances.py}
\end{figure}

To illustrate the accuracy of the numerical procedure for deriving the model covariance matrix, $\bm{\Sigma^\prime}$, we present in \cref{fig:covariances} the MSD covariance matrix for \num{4096} random-walk simulations, as described above, at three differing levels of approximation:
\cref{fig:covariances}a shows the numerically converged covariance matrix, $\bm{\Sigma}$, computed using the data from all \num{4096} simulations (\cref{fig:covariances}a); \cref{fig:covariances}b shows the corresponding analytical covariance matrix, $\mathbf{\Sigma^\prime}$, as defined by \cref{equ:cvv} and parametrised using analytical long-time-limit variances $\var{\oMSDi}$; and \cref{fig:covariances}c shows the average estimated matrix obtained by parametrising \cref{equ:cvv} using variances estimated from a single simulation trajectory, then reconditioning, with the average taken over all \num{4096} matrices obtained from the \num{4096} input simulations. 

\begin{figure*}[ht!]
    \centering
    \resizebox{\textwidth}{!}{\includegraphics*{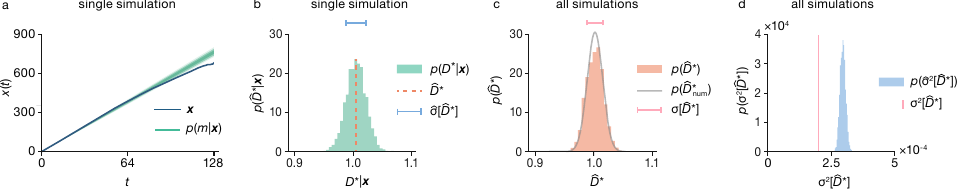}}
    \caption{
        (a) Observed MSD from a single simulation of 128 particles undergoing a 3D-lattice random walk of 128 steps per particle (dark line).
        The green shading shows the corresponding posterior distribution $\prob{\model|\oMSD}$ of linear models compatible with the observed MSD data $\oMSD$, calculated using the scheme described in the main text.
        The variegated shading indicates compatibility intervals of \SIlist[list-final-separator = {, and }]{1;2;3}{\sigma}$[\prob{\model|\oMSD}]$.
        (b) The marginal posterior distribution $\prob{\Dest|\oMSD}$ obtained from the posterior distribution of linear models in (a).
        The mean of this distribution gives the point estimate $\Dest$ for this simulation input data.
        The blue horizontal bar shows an interval of one standard deviation in $\prob{\Dest|\oMSD}$.
        (c) Probability distribution of point-estimates $\prob{\Dest}$ obtained from \num{4096} individual random-walk simulations.
        Each simulation has been analysed as in (a) and (b) to yield a single corresponding point estimate $\Dest$.
        The grey line shows the distribution of point estimates, $\prob{\Dest_\mathrm{num}}$, obtained using Bayesian regression with a mean vector and numerical covariance matrix derived from the complete dataset of all \num{4096} simulations.
        The pink horizontal bar shows an interval of one standard deviation in $\prob{\Dest}$.
        (d) Probability distribution of estimated variances, $\varest{\Dest}$, for individual random-walk simulations, compared to the true sample variance (pink vertical line) $\var{\Dest}$.
    } 
    \label{fig:random_walk}
    \script{random_walk.py}
\end{figure*}

While the analytical and average estimated covariance matrices show some systematic deviation from the numerically converged covariance matrix, the general correlation structure is preserved.
The discrepancy between the model and numerical covariance matrices largely stems from the approximation made in deriving the analytical form that $t$ is large, which leads to an overestimation of the variance at low $t$.
Despite this, the average estimated covariance matrix reproduces well the correlation structure of the true numerical covariance matrix, and, as we show below, the covariance matrices estimated from individual simulation trajectories can be used within approximate GLS or Bayesian regression schemes to estimate $\D$ and $\var{\Dest}$.

\section{Validation}

To demonstrate the complete approximate Bayesian regression scheme we present two example use-cases.
First, we consider a simple 3D-lattice random walk; in this case the true self-diffusion coefficient $\D$ is specified by the simulation parameters and a well-converged numerical covariance matrix can be obtained at relatively low computational cost, allowing us to directly compare the estimates produced by our method to ``best case'' estimates from a hypothetical method with access to the true covariance matrix.
Second, we consider an example real-world system, the lithium-ion solid electrolyte \ce{Li7La3Zr2O12} (LLZO), which represents an application of our method to a well-studied material of practical interest for solid-state lithium-ion batteries \cite{MuruganEtAl_AngewChemIntEd2007,burbano_sparse_2016,morgan_lattice_2017,SquiresEtAl_PhysRevMater2022}.

\Cref{fig:random_walk}a shows the observed MSD from a single 3D-lattice random-walk simulation, along with the estimated posterior distribution of linear models compatible with the observed MSD data, $\prob{\model|\oMSD}$, calculated via \cref{equ:bayes,equ:loglike}.
The corresponding marginal posterior distribution of estimated diffusion coefficients $\prob{\D|\oMSD}$ is shown in \cref{fig:random_walk}b; this distribution is approximately Gaussian and is centred close to the true self-diffusion coefficient $\D = \num{1}$, demonstrating that for this example trajectory we obtain a good point-estimate of $\D$.

To evaluate the overall performance of our method, we repeat our analysis on the full set of \num{4096} random-walk simulations.
\cref{fig:random_walk}c presents a histogram of the resulting point estimates of $\D$, with each estimate derived as the mean of the posterior distribution $\prob{\D|\oMSD}$ using input data from each individual simulation.
We also show the probability distribution of estimated diffusion coefficients obtained using Bayesian regression with a mean vector and covariance matrix derived numerically from all \num{4096} simulations (solid line).
This latter distribution represents the distribution of ``best possible'' estimates of $\D$ and exhibits the minimum possible theoretical variance.
The close agreement between these two distributions demonstrates that our approximate Bayesian regression scheme yields nearly optimal estimates of $\D$ using data from individual simulations.
The distribution of estimated diffusion coefficients from single simulations is slightly broader than the exact numerical results.
This minor deviation is a consequence of the noise present in data obtained from a single simulation trajectory.

We next consider the degree to which our method can quantify the uncertainty in $\Dest$ when using input data from a single simulation.
\cref{fig:random_walk}d shows the distribution of estimated variances $\varest{\Dest}$, with each sample calculated from an individual simulation trajectory.
We also show the true variance of individual point estimates, $\var{\Dest}$, which characterises the spread of the histogram in \cref{fig:random_walk}c.
The distribution of estimated variances is biased relative to the true variance, due to numerical differences between the true covariance matrix $\mathbf{\Sigma}$ and the estimated covariance matrix $\mathbf{\Sigma^\prime}$ (further details are provided in the SI).
In general, however, the distribution of the estimated variance shows good agreement with the true sample variance.
Notably, the precision of this estimate is significantly greater than obtained using OLS or WLS and their respective textbook statistical formulae.

We also benchmark our method using data from simulations of the lithium-ion solid electrolyte, cubic \ce{Li7La3Zr2O12} (c-LLZO).
We performed a single simulation of \num{1536} atoms (\num{448} Li ions) at \SI{700}{K} for \SI{6}{\nano\second} (full simulation details are given in the Methods section).
To generate multiple statistically equivalent trajectories, we partitioned the output simulation data into \num{192} effective trajectories, each $\sim\SI{500}{\ps}$ in length, and containing data for \num{28} lithium ions, which were selected randomly from the complete set of \num{448} lithium ions without replacement.
We then performed approximate Bayesian regression, as above, on each effective trajectory, excluding the first \SI{10}{ps} of MSD data in each case to remove short-time data corresponding to the ballistic and sub-diffusive regimes \cite{burbano_sparse_2016,he_statistical_2018}.

\begin{figure}[tb]
    \centering
    \resizebox{\columnwidth}{!}{\includegraphics*{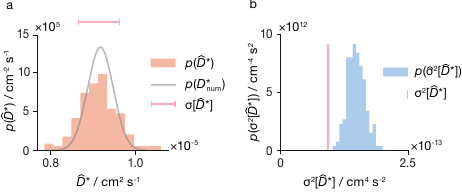}}
    \caption{
      (a) Probability distribution of point estimates $\prob{\Dest}$ for \num{192} effective simulations of LLZO (orange histogram).
      The grey line shows the distribution $\prob{\Dest_\mathrm{num}}$ obtained using Bayesian regression with the complete LLZO dataset as input.
      The pink bar shows an interval of one standard deviation $\sigma[\prob{\Dest}]$.
      (b) Probability distribution of estimated variances, $\varest{\Dest}$, for individual LLZO effective simulations, compared to the true sample variance (pink vertical line) $\var{\Dest}$.
    }
    \label{fig:diffusion}
    \script{diffusion.py}
\end{figure}

The resulting distribution of the point estimates, $\Dest$, from analysis of all \num{192} effective trajectories, is shown in \cref{fig:diffusion}a.
As above, we also show the corresponding distribution of $\Dest$ estimates obtained via Bayesian regression using a well-converged numerical covariance matrix calculated from the full LLZO dataset.
The distribution $\prob{\Dest}$ obtained using the model covariance matrix and parametrised separately for each individual effective simulation is highly similar to that obtained using the aggregate numerical covariance matrix calculated from the complete simulation dataset.
This close agreement mirrors the results for our random walk simulations, and confirms that our method yields accurate and statistically efficient estimates for $\D$, even for real-world simulation data.

We also consider the probability distribution of estimates of the variance in $\Dest$ calculated for each effective trajectory (\cref{fig:diffusion}b), which we compare to the true variance in $\Dest$ for our method; i.e., the variance of the histogram in \cref{fig:diffusion}a.
While the estimated variances deviate somewhat from the true distribution $\prob{\var{\Dest}}$, the agreement is reasonable and mirrors our results for the random walk simulations.
Hence, our method provides reasonably accurate estimates of the uncertainty in $\Dest$ for our c-LLZO dataset, even when applied to single effective trajectories with limited displacement data (only 28 mobile ions, and \SI{500}{ps} simulation length).

\section{$\var{\Dest}$ scaling and comparison to OLS, WLS, and GLS}

\begin{figure}[htb]
    \centering
    \resizebox{\columnwidth}{!}{\includegraphics*{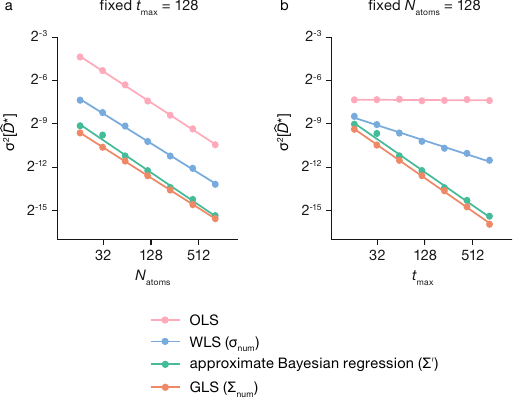}}
    \caption{
      Scaling of $\var{\Dest}$ with simulation size for OLS (pink), WLS (blue), our approximate Bayesian regression method (green), and GLS (orange).
      (a) Scaling versus number of mobile particles, $N_\mathrm{atoms}$.
      (b) Scaling versus total simulation time, $t_\mathrm{max}$.
      Solid lines show fitted power law relationships for each dataset.
      The WLS and GLS data are obtained using numerically determined variances and covariance, respectively, from a set of \num{512} repeat simulations for each combination of $N_\mathrm{atoms}$ and $t_\mathrm{max}$.
    }
    \label{fig:stat_eff}
    \script{stat_eff.py}
\end{figure}

\Cref{fig:stat_eff} presents an analysis of the variation in $\var{\Dest}$ as the number of mobile particles (\cref{fig:stat_eff}a) and the total simulation time (number of steps) (\cref{fig:stat_eff}b) are changed.
We compare four methods for estimating $\D$ from the observed MSD data: OLS, WLS, the approximate Bayesian regression method described here, and GLS.
When estimating $\D$ using WLS and GLS, we calculate the variances and the covariance matrix, respectively, numerically, using the complete set of \num{512} simulations.
Each data point in \cref{fig:stat_eff} represents the variance across point-estimates of $\D$ derived from \num{512} individual 3D-lattice random walk simulations, for each combination of $N_\mathrm{atoms}$ and $t_\mathrm{max}$.
The GLS dataset corresponds to an optimally efficient estimator for linear regression of observed MSD data, and is equivalent to performing Bayesian regression with the converged numerical covariance matrix and an uninformative prior.

Our approximate Bayesian regression method performs similarly to GLS with a numerically converged covariance matrix, and gives significantly reduced uncertainty in $\Dest$ compared to OLS or WLS, for all simulation sizes and lengths considered.
Moreover, our method scales better than OLS or WLS as the total simulation time is increased.
This approximate Bayesian regression method therefore presents a significant improvement over more conventional methods such as OLS and WLS, by enabling more precise estimates of $\D$ across varied simulation sizes at equivalent computational cost.

\section{Summary and Discussion}

We have introduced and demonstrated an approximate Bayesian regression method for estimating the self-diffusion coefficient, $\D$, from molecular dynamics simulation data.
We consider the observed mean-squared displacement data from a single simulation as a random sample, $\bm{X}$, from a population of potential MSDs generated by equivalent replica simulations, $\bm{X}\sim p(\oMSD)$.
We model this population using a multivariate normal distribution, $p(\oMSD) = \mathcal{N}(\model, \mathbf{\Sigma})$, with mean vector $\model = 6\D\bm{t} + c$, where $\D$ and $c$ are model parameters to be determined.

To model the covariance matrix, we use an analytical solution derived for an equivalent system of freely diffusing particles.
To parameterise this model covariance matrix, we renormalise the variance of the observed squared displacements from the input simulation trajectory, followed by a reconditioning step to ensure a positive-definite matrix.
The resulting model covariance matrix preserves the correlation structure of the true simulation MSD covariance matrix, and gives a multivariate normal model for the population of observable simulation MSDs that depends solely on the model parameters, $\D$ and $c$.

We use Markov-Chain Monte Carlo to sample the posterior distribution of linear models compatible with the observed MSD data.
This approach yields a marginal posterior distribution, $\prob{\D | \oMSD}$, that gives a statistically efficient point estimate for $\D$ and allows the associated statistical uncertainty, $\var{\Dest}$, to be quantified.

We have benchmarked our approach using simulation data for an ideal 3D lattice random walk and for the lithium-ion solid electrolyte \ce{Li7La3Zr2O12} (LLZO).
In both cases, we obtain a distribution of estimates for $\D$ that closely matches the theoretically optimal distribution obtained using a well-converged numerical covariance matrix derived from a large number of replica simulation trajectories.

We obtain estimates for $\D$ that are unbiased, with near-optimal statistical efficiency, using input data from single simulation trajectories.
The approximate Bayesian regression scheme therefore provides more accurate single-point estimates of the self-diffusion coefficient than the commonly used OLS or WLS methods, when applied to the same input simulation data.
The improved statistical efficiency of this method, when compared to OLS or WLS, enables the estimation of $\D$ with equivalent accuracy from considerably smaller simulations---either in terms of timescale or system size. 
This reduces the overall computational cost when compared to studies that use OLS or WLS for estimating a linear fit to simulation MSD data.
Alternatively, this approach enables the estimation of $\D$ with greater precision, given simulation trajectories of equal size.

Our method also provides reasonable estimates of the statistical uncertainty in the estimated value $\Dest$, in contrast to OLS and WLS which systematically significantly underestimate the uncertainty in regression coefficients when applied to simulated MSD data. 
While these estimated statistical uncertainties can still differ from the true (but unknown) uncertainty in $\Dest$, particularly when using short-timescale simulation data, they allow for scientifically meaningful comparisons to be made between estimated diffusion coefficients across different systems or under varying conditions, such as changes in temperature, or between computational findings and experimental results.
Furthermore, these uncertainties allow for quantitative downstream analysis, such as the application of Arrhenius (on non-Arrhenius) type models to describe the temperature dependence of self-diffusion.

The approximate Bayesian regression scheme presented here provides a statistically efficient means of estimating the self-diffusion coefficient, $\D$, from molecular dynamics simulation data.
It improves upon textbook approaches by providing accurate point estimates of $\D$ with near-optimal statistical efficiency, while also providing a reasonable description of the uncertainty in these estimates.
The high statistical efficiency of our method allows for the use of smaller simulations, which can significantly reduce computational costs.
Overall, our method offers significant advantages over more conventional methods of estimating self-diffusion coefficients from atomistic simulations.
We have implemented this procedure in the open-source package \textsc{kinisi} \cite{mccluskey_kinisi_2022}, which we hope will support its use within the broader simulation community.

\section{Methods}

\subsection{Numerical implementation in KINISI}
\label{sec:implementation}
\textsc{kinisi}-1.1.0 was used for all analyses presented in this work.  

When calculating the observed mean squared displacement at each time interval $t$ (see \cref{equ:observed_msd}), \textsc{kinisi} uses overlapping sliding window sampling.
For a given time interval, $t$, the maximum number of observations is $N_{\mathrm{atoms}} \times (N_{t} - i)$ displacements, where $N_{\mathrm{atoms}}$ is the number of mobile atoms, $N_{t}$ is the total number of timesteps, and $i$ is the index of the timestep (where \num{1} is the index for the shortest timestep).
To estimate the variance of the observed MSD, we rescale the variance of observed squared displacements by the number of numerically-independent sub-trajectories in the simulation, $\nind{i} = N_{\mathrm{atoms}} \times N_{t} / i$, as presented in Eqn.~\ref{equ:varestMSD}.  

The parametrisation of the covariance matrix from the variances $\var{\oMSDi}$ and the number of independent observations $\nind{i}$ is defined by Eqn.~\ref{equ:cvv}. 
The covariance matrix is only constructed for values of $t$ where the particle motion is considered to be in the long-time diffusive limit.
In practice, this threshold is set by the user to a value appropriate for their system and simulation data. 
For the examples presented in the main manuscript, we consider particles to be in the diffusive regime from $t=\num{2}$ for the random walk trajectories and from $t=\SI{10}{\pico\second}$ for the LLZO simulations.
The covariance matrix is reconditioned using the minimum eigenvalue method~\cite{TabeartEtAl_TellusDynMeteorolOceanogr2020}, with a maximum condition number of \num{1e16} for all simulations.

To estimate $\Dest$ from a given set of MSD data, \textsc{kinisi} uses ordinary least squares to obtain an initial guess for the gradient and intercept of the linear model that best describes the observed MSD.
This initial guess is then used as the starting point for minimising the negative maximum a posteriori (the peak of the posterior distribution as per \cref{equ:bayes}), with the improper prior that $\D \ge 0$~\cite{broyden_convergence_1970,fletcher_new_1970,goldfarb_family_1970,shanno_conditioning_1970}.
We note that the Bayesian regression formalism presented here allows for the use of alternative informative priors in cases where the user has some prior knowledge of the system being simulated that they wish to incorporate into their analysis.
The log-likelihood calculation (\cref{equ:loglike}) uses the Moore--Penrose generalisation of the inverse of a Hermitian matrix~\cite{moore_on_1920,bjerhammar_application_1951,penrose_generlized_1955}.

To sample the joint posterior probability distribution of the linear model, \textsc{kinisi} uses the \textsc{emcee} package~\cite{foremanmackey_emcee_2019}, which implements Goodman and Weare's affine invariant Markov chain Monte Carlo ensemble sampler~\cite{goodman_ensemble_2010}. 
When sampling $\prob{\D|\model}$ we again apply the improper prior $\D \ge 0$.
The sampling process uses \num{32} walkers for \num{1500} steps, with the first \num{500} steps discarded as a burn-in period.
The sampled chains are thinned such that only every 10th value is retained, yielding \num{3200} points sampled from the posterior distribution $\prob{\D|\model}$.
These points can then be plotted as a histogram (as in \cref{fig:random_walk}b), and summary statistics $\Dest$ and $\varest{\Dest}$ can be derived.

\subsection{LLZO simulations}
Classical molecular dynamics simulations were run using the \textsc{metalwalls} code~\cite{marin_metalwalls_2020}. 
We used the DIPPIM polarisable ion force field, as parameterised by Burbano \emph{et al.}~\cite{burbano_sparse_2016}. 
We simulated the cubic phase of LLZO in the NVT ensemble at a temperature of \SI{700}{\kelvin}.
Simulations were run for \SI{6}{\nano\second} with a \SI{0.5}{\femto\second} timestep. 
To control temperature, we used a Nos\'{e}-Hoover thermostat, with a relaxation time of \SI{121}{\femto\second} (5000 $\hbar / E_{h}$)~\cite{nose_unified_1984,hoover_canonical_1985,martyna_nose_1992}. 
Simulations were performed using $2 \times 2 \times 2$ supercells with \SI{1536}{atoms} following the same protocol as in Ref.~\citenum{burbano_sparse_2016}.

\section*{Supporting Information}

Derivation of long-time limit covariance matrix for a system of freely diffusing particles, comparison of variance rescaling (\cref{equ:varestMSD}) and block renormalisation approaches, discussion of bias in the distribution of estimated variance of the estimated diffusion coefficient, and a comparison of OLS, WLS, and GLS as estimators for $\D$ applied to MSD data from simulations of \ce{Li7La3Zr2O12} (LLZO).
Additional Electronic Supplementary Information (ESI) available at Ref.~\cite{mccluskey_github_2022} under an MIT license: A complete set of analysis/plotting scripts allowing for a fully reproducible and automated analysis workflow, using \textsc{showyourwork}~\cite{luger_showyourwork_2021}.
The LLZO raw simulation trajectories are available on Zenodo shared under a CC BY-SA 4.0 licence~\cite{coles_llzo_zenodo_2022}.
The method outlined in this work is implemented in the open-source Python package \textsc{kinisi}~\cite{mccluskey_kinisi_2022}, which is available under an MIT license.

\section*{CR\lowercase{e}d\lowercase{i}T author statement}

A.R.M.: Conceptualization, Formal Analysis, Investigation, Methodology, Software, Visualisation, Writing---original draft.
S.W.C.: Methodology, Resources, Writing---review and editing.
B.J.M.: Conceptualization, Methodology, Software, Writing---review and editing.

\section*{Acknowledgements}

The authors thank Jacob M.\ Dean and Tim Rogers for their valuable input in checking the mathematical derivations that make up the appendices, the beta-testers for the \textsc{kinisi} package, and Rodrigo Luger and Daniel Foreman-Mackey for their help using \textsc{showyourwork}~\cite{luger_showyourwork_2021}.
This work used the Isambard 2 UK National Tier-2 HPC Service (http://gw4.ac.uk/isambard/) operated by GW4 and the UK Met Office, and funded by EPSRC (EP/T022078/1).
The authors acknowledge the University of Bath's Research Computing Group for their support in running the LLZO molecular dynamics simulations.
Other simulations and analyses were carried out using the Data Management and Software Centre computing cluster at the European Spallation Source ERIC.
S.W.C. and B.J.M. acknowledge the support of the Faraday Institution through the CATMAT project (grant number FIRG016). 
B.J.M. acknowledges support from the Royal Society (UF130329 and URF\textbackslash R\textbackslash 191006). 

\bibliographystyle{naturemag}
\bibliography{bib}

\let\addcontentsline\oldaddcontentsline

 \onecolumngrid
 \clearpage 
 \twocolumngrid

 \appendix
 \renewcommand\thesection{S-\Roman{section}}
 \counterwithout{figure}{section}
 \renewcommand\thefigure{S-\arabic{figure}}
 \setcounter{figure}{0}
 \counterwithout{equation}{section}
 \renewcommand\theequation{S-\arabic{equation}}
 \setcounter{equation}{0}
 \counterwithout{table}{section}
 \renewcommand\thetable{S-\arabic{table}}
 \setcounter{table}{0}
 \pagenumbering{arabic} 
 \renewcommand\thepage{S-\arabic{page}}
 \addtocontents{toc}{\protect\setcounter{tocdepth}{0}}

 \title{Supplemental Material for ``\papertitle''}
 \maketitle
 This document presents supplementary material for the manuscript ``\papertitle''.
It contains the following sections:
\renewcommand{\labelitemi}{}
\begin{itemize}
    \item \ref{sec:ran}. The derivation of the covariance matrix in the long-time limit for freely diffusion particles.
    \item \ref{sec:var_est}. Further details of the variance rescaling method for estimating $\var{\oMSDi}$, described in the main text, and a comparison to the block renormalisation method for estimating the variance of the mean for serially-correlated data of Flyvbjerg and Petersen~\cite{FlyvbjergAndPetersen_JChemPhys1989}.
    \item \ref{sec:bias}. Discussion of the origin of bias in the distribution of the estimated variance of the estimated diffusion coefficient, $\prob{\varest{\Dest}}$.
    \item \ref{sec:llzo}. A comparison of OLS, WLS, and GLS as estimators for $\D$ applied to MSD data from simulations of \ce{Li7La3Zr2O12} (LLZO).
\end{itemize}
A repository containing the analysis and plotting code used to generate all results and figures in the main manuscript and this supplemental material document is available at \url{www.github.com/arm61/msd-errors}~\cite{mccluskey_github_2022}, under MIT (code) and CC BY-SA 4.0 (figures and text) licenses.
This repository includes a fully reproducible \code{showyourwork} workflow, which allows complete reproduction of the analysis, plotting of figures and compilation of the manuscript{}s.
The corresponding LLZO simulation datasets are openly available under the CC BY-SA 4.0 licence \cite{coles_llzo_zenodo_2022}. 

\twocolumngrid

\section{Derivation of the long-time limit covariance matrix for a system of freely diffusing particles.}
\label{sec:ran}
In the main manuscript we present the result that the covariance matrix for a system of freely diffusing particles, in the long-time limit, has the form
\begin{equation}
  \Sigma^\prime\left[\oMSDi, \oMSDj\right]= \Sigma^\prime\left[\oMSDj, \oMSDi\right] =
  \var{\oMSDi} \frac{\nind{i}}{\nind{j}},\hspace{1em} \forall\,i \leq j,
  \label{equ:cvv_SI}
\end{equation}
where $\oMSDi$ is the observed mean-squared displacement (MSD) for time interval $i$ and $\nind{i}$ is the number of statistically independent observed squared displacements averaged over to compute the mean value.

To derive this result, we first present a derivation of the expected variance for the MSD at timestep $i$, $\var{\oMSD}$, following the approach of Smith and Gillan~\cite{smith_random_1996}.
We then derive an expression for the covariance $\Sigma^\prime\left[\oMSDi, \oMSDj\right]$ to obtain the result above.

For a single particle undergoing a one-dimensional random walk with step size $\kappa$, each step gives a displacement $h = \pm \kappa$.
After $n$ steps, the MSD, $\oMSDn$, is given by
\begin{equation}
  \begin{aligned}
    \oMSDn &= \left[\sum_i^n h_i\right]^2\\
           &= \sum_i^n\sum_j^n h_i h_j \\
           &= \sum_i^n h_i^2 + \sum_i^n\sum_{j\neq i}^n h_ih_j.
  \end{aligned}
\end{equation}
The expected MSD in the long-time limit, $\mathbb{E}(\oMSDn) = \moMSDn$, is obtained by averaging over all permutations of $h_i$ and $h_j$:
\begin{equation}
  \begin{aligned}
    \moMSDn &= \sum_i^n \left<h_i^2\right> + \sum_i^n\sum_{j\neq i}^n\left<h_i h_j\right>.
  \end{aligned}
\end{equation}
For a random walk, the second term averages to zero for all $h_i$ and $h_j$, and
\begin{equation}
  \begin{aligned}
    \moMSDn &= \sum_i^n \left<h_i^2\right> \\
            &= n\kappa^2.
  \end{aligned}
  \label{equ:msd_SI}
\end{equation}
Hence the expected value for the mean-squared displacement increases linearly with the number of steps taken.

The variance in the observed MSD, $\var{\oMSDn}$, is given by the standard statistical formula
\begin{equation}
    \var{\oMSDn} = \left<{\left[\oMSDn - \moMSDn\right]}^2\right>,
\end{equation}
which can be expanded as
\begin{equation}
    \begin{aligned}
        \var{\oMSDn} &= \left<\oMSDn^2\right> - 2\moMSDn\left<\oMSDn\right> + \moMSDn^2,\\
                     &= \left<\oMSDn^2\right> - \moMSDn^{\,2}.
    \end{aligned}
\end{equation}
The first term can be expanded in terms of displacements $h$ as
\begin{equation}
  \left<\oMSDn^2\right> = \left<\sum_i^n\sum_j^n\sum_k^n\sum_l^n h_i h_j h_k h_l\right>,
  \label{equ:big_av}
\end{equation}
which can be simplified by noting that $h_i$, $h_j$, $h_k$, and $h_l$ are uncorrelated when $i \neq j \neq k \neq l$, and the only terms that contribute to the average are those where $h_ih_jh_kh_l$ is guaranteed to be non-zero:
\begin{itemize}
    \item[(a)] $i = j = k = l$;
    \item[(b)] $(i = j) \neq (k = l)$;
    \item[(c)] $(i = k) \neq (j = l)$;
    \item[(d)] $(i = l) \neq (j = k)$. 
\end{itemize}
From (a) we obtain
\begin{equation}
    \left<\sum_i^nh_i^4\right> = n\kappa^4,
\end{equation}
and from (b), (c), and (d), which are equivalent, we obtain
\begin{equation}
    \left<\sum_i^n\sum_j^nh_i^2h_j^2\right> = (n\kappa^2)^2 = n^2\kappa^4.
\end{equation}
This gives
\begin{equation}
    \left<\oMSDn^2\right> = (3n^2 + n)\kappa^4,
\end{equation}
which, in the limit $n \to \infty$, approaches
\begin{equation}
    \left<\oMSDn^2\right> = 3n^2\kappa^4.
    \label{equ:infty}
\end{equation}
Combining this result with \cref{equ:msd_SI}, we can express the variance in the mean-squared displacement as
\begin{equation}
    \var{\oMSDn} = 3n^2\kappa^4 - n^2\kappa^4 = 2n^2\kappa^4,
    \label{equ:varmsd}
\end{equation}
i.e., $\var{\oMSDn}$ increases quadratically with the number of steps taken, or, equivalently, with time.

Equation~\ref{equ:varmsd} gives the variance of the mean squared displacement for a single particle considering a single time-origin.
We can obtain improved statistics by averaging over statistically equivalent observed squared displacements (see Eqn.~2 in the main text), which can be achieved by averaging over mobile particles or by averaging over time origins.
This averaging over equivalent observations reduces the variance in the observed MSD to
\begin{equation}
    \var{\oMSDn} = \frac{2n^2\kappa^4}{\nind{n}},
\label{equ:der_var}
\end{equation}
where $\nind{n}$ is the total number of statistically independent squared displacements that contribute to $\oMSDi$.
In the long-time limit, $\nind{n}$ is given by the product of the number of mobile particles and the number of non-overlapping time-windows of length $i$ in our simulation trajectory.
Note that $\nind{n}$ considers non-overlapping time windows, since mutually overlapping time-windows give correlated squared displacements.

The results for a one-dimensional lattice above (\cref{equ:msd_SI,equ:der_var}) can be extended to a $d$-dimensional lattice, to give
\begin{equation}
    \moMSDn_{d} = \sum^d{\frac{n\kappa^2}{d}} = n\kappa^2,
\end{equation}
with variance
\begin{equation}
    \var{\oMSDn}_d = \sum^d{\frac{2n^2\kappa^4}{d^2\nind{n}}} = \frac{2n^2\kappa^4}{d\nind{n}},
\end{equation}
Because each step is equally likely to move a particle along each of the $d$ dimensions, the term $n$ in \cref{equ:msd_SI,equ:der_var} is replaced here with $n/d$.

The analysis above can be extended to consider the covariance between two different numbers of steps, $n$ and $n+m$, in the random walk where the expected MSDs will be
\begin{equation}
    \begin{aligned}
        \moMSDn &= n\kappa^2; \\
        \moMSDm &= (n+m)\kappa^2.
    \end{aligned}
\end{equation}
The covariance between these is defined as
\begin{equation}
    \Sigma \left[\oMSDn, \oMSDm \right] = \left<{\left[{\oMSDn - \moMSDn}\right] \left[{\oMSDm - \moMSDm}\right]}\right>, 
\end{equation}
which can be expanded as
\begin{equation}
    \begin{aligned}
        \Sigma \left[\oMSDn, \oMSDm \right] = \big\langle & \oMSDn\oMSDm - \oMSDn\moMSDm \\
        & - \moMSDn\oMSDm + \moMSDn\moMSDm\big\rangle,
    \end{aligned}
\end{equation}
and then reformulated to give
\begin{equation}
    \Sigma \left[\oMSDn, \oMSDm \right] = \left<\oMSDn\oMSDm\right> - \moMSDn\moMSDm,
    \label{equ:pair}
\end{equation}
where 
\begin{equation}
    \begin{aligned}
        \moMSDn\moMSDm &= \oMSDn\oMSDm \\
                       &= n\kappa^2(n+m)\kappa^2 \\
                       &= n(n+m)\kappa^4
    \end{aligned}
\end{equation}
and, by analogy to \cref{equ:big_av},
\begin{equation}
    \left<\oMSDn\oMSDm\right> = \left<\sum_i^n\sum_j^n\sum_k^{n+m}\sum_l^{n+m} h_i h_j h_k h_l\right>,
\end{equation}
which we can rewrite as
\begin{equation}
    \begin{aligned}
        \left<\oMSDn\oMSDm\right> = \Bigg\langle & \sum_{i=1}^n\sum_{j=1}^n\sum_{k=1}^{n}\sum_{l=1}^{n} h_i h_j h_k h_l \\
        & + \sum_{i=1}^n\sum_{j=1}^n\sum_{k=1}^{n}\sum_{l=n+1}^{n+m} h_i h_j h_k h_l \\
        & + \sum_{i=1}^n\sum_{j=1}^n\sum_{k=n+1}^{n+m}\sum_{l=1}^{n} h_i h_j h_k h_l \\
        & + \sum_{i=1}^n\sum_{j=1}^n\sum_{k=n+1}^{n+m}\sum_{l=n+1}^{n+m} h_i h_j h_k h_l \Bigg\rangle.
    \end{aligned}
    \label{equ:vbig}
\end{equation}
The second and third terms in \cref{equ:vbig} tend to zero as there is an equal probability of positive and negative displacements. 
This reduces \cref{equ:vbig} to 
\begin{equation}
    \begin{aligned}
        \left<\oMSDn\oMSDm\right> = & \Bigg\langle \sum_{i=1}^n\sum_{j=1}^n\sum_{k=1}^{n}\sum_{l=1}^{n} h_i h_j h_k h_l \Bigg\rangle \\
        & + \Bigg\langle \sum_{i=1}^n\sum_{j=1}^n\sum_{k=n+1}^{n+m}\sum_{l=n+1}^{n+m} h_i h_j h_k h_l \Bigg\rangle, 
    \end{aligned}
\end{equation}
and using \cref{equ:infty} gives
\begin{equation}
    \left<\oMSDn\oMSDm\right> = 3n^2\kappa^4 + \Bigg\langle \sum_{i=1}^n\sum_{j=1}^n\sum_{k=n+1}^{n+m}\sum_{l=n+1}^{n+m} h_i h_j h_k h_l \Bigg\rangle.
\end{equation}
We can rewrite this as
\begin{equation}
    \left<\oMSDn\oMSDm\right> = 3n^2\kappa^4 + \Bigg\langle \sum_{i=1}^n\sum_{j=1}^n h_i h_j \Bigg\rangle \Bigg\langle \sum_{k=n+1}^{n+m}\sum_{l=n+1}^{n+m} h_k h_l \Bigg\rangle,
\end{equation}
where the following holds,
\begin{equation}
    \begin{aligned}
        \left<\oMSDn\oMSDm\right> &= 3n^2\kappa^4 + n\kappa^2 m\kappa^2 \\
                                  &= 3n\kappa^4 + nm\kappa^4.
    \end{aligned}
\end{equation}
Putting this result into \cref{equ:pair} allows the covariance to be written as
\begin{equation}
    \begin{aligned}
    \Sigma^\prime \left[\oMSDn, \oMSDm \right] &= 3n^2\kappa^4 + nm\kappa^4 - n(n+m)\kappa^4 \\
                                        &= 3n^2\kappa^4 - n^2\kappa^4 = 2n^2\kappa^4,
    \end{aligned}
    \label{equ:cov_der}
\end{equation}
where we use the $\Sigma^\prime$ notation to identify that this is in the long-time limit. 

In this case, the covariance depends only on the number of overlapping points, $n$, between the two time intervals. 
We can rationalise this by noting that for a random walk any non-overlapping points will be completely uncorrelated and therefore have a covariance of \num{0}. 
Similar to the case for the variance, the covariance derived in \cref{equ:cov_der} is that for a single particle at a single time origin. 
The number of independent observed squared displacements for a given covariance should be the minimum number of shared independent observed squared displacements between the two time intervals, which is $\nind{n+m}$. 
Therefore, the covariance, scaled by the number of contributing independent observations, in the long-time limit, is
\begin{equation}
    \Sigma^\prime \left[\oMSDn, \oMSDm \right] = \frac{2n^2\kappa^4}{\nind{n+m}}.
\end{equation}
Similar to the MSD and the variance, the covariance can be written for $d$-dimensions as 
\begin{equation}
    \Sigma^\prime \left[\oMSDn, \oMSDm \right] = \frac{2n^2\kappa^4}{d\nind{n+m}}.
\end{equation}
The covariance can be calculated directly from the variance by recognising that both depend on the number of overlapping points, $n$, as follows
\begin{equation}
    \Sigma^\prime \left[\oMSDn, \oMSDm \right] = \var{\oMSDn}\frac{\nind{n}}{\nind{n+m}}.
\end{equation}
This is then rewritten in terms of $i$ and $j$ to give, \cref{equ:cvv_SI}.

Using the equivalence of $2d\D t \equiv n\kappa^2$~\cite{howard_reports_1964}, \cref{equ:msd_SI,equ:cvv_SI} can be rewritten in terms of $t$ (or $t_1$ and $t_2$) and the diffusion coefficient, for any dimensionality of lattice random walk,
\begin{equation}
    \oMSDs{t} = 2d \D t,
    \label{equ:der_msd_fick}
\end{equation}
and 
\begin{equation}
    \Sigma^\prime \left[\oMSDs{t_1}, \oMSDs{t_2} \right] = 8d{(\D)}^2 t_1^2\frac{N^\prime(t_2)}{N^\prime(t_1)},\hspace{1em} \forall\,t_1 \leq t_2.
\end{equation}

\section{Estimating $\var{\oMSDi}$: variance rescaling versus block renormalisation}
\label{sec:var_est}
The approximate Bayesian regression scheme described in our main manuscript uses a model covariance matrix parametrised by the variance of the observed MSD as a function of time, denoted $\var{\oMSDi}$. 
Generally, $\var{\oMSDi}$ is unknown and must be estimated from the input simulation data.

\begin{figure}[tb]
  \centering
  \resizebox{9cm}{!}{\includegraphics*{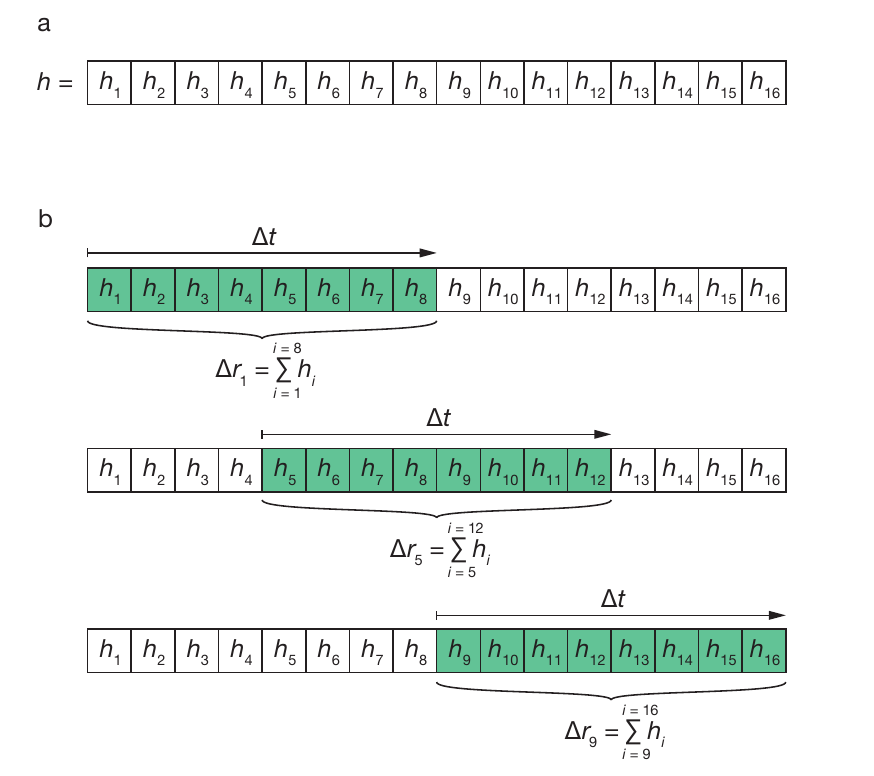}}  
    \caption{\label{fig:displacements}Schematic showing the construction of three statistically-equivalent sub-trajectories of length $\Delta t = 8$ from a full simulation trajectory. (a) The full trajectory consists of a series of displacements, $h_i$, from each simulation timestep. (b) Three exemplar sub-trajectories of length $\Delta t = 8$. Even in the case of a random walk (i.e., $h=\pm\kappa$), $\Delta r_1$ and $\Delta r_5$ are necessarily correlated, because both summations include $[h_5 + h_6 + h_7 + h_8]$. $\Delta r_5$ and $\Delta r_9$ are also obtained from mutually overlapping time windows and are therefore also correlated. For a random walk, $\Delta r_1$ and $\Delta r_9$ are uncorrelated. For a non-random (correlated) walk, non-overlapping time windows are uncorrelated in the limit that the window length, $\Delta t$, is much greater than the correlation time in $h$, which is the case if we are in the long-time linear region of the MSD where Eqn.~1 is valid.\footnote{The result that displacements calculated from overlapping time windows are correlated has been reported as an empirical result for simulations of a Lennard--Jones fluid in Ref.~\onlinecite{PranamiAndLamm_JChemTheoryComput2015}.}} 
\end{figure}

In the main manuscript, we describe an estimation approach that involves rescaling the observed variance of the squared displacement for time interval $i$ by the number of numerically-independent contributing sub-trajectories, $\nind{i}$ (Eqn.~7). 
We define a sub-trajectory as the sequence of displacements of one particle over a time interval of length $i$, and consider sub-trajectories to be numerically independent under two conditions: either when they describe displacements of different particles, or when they describe displacements of the same particle but are calculated from non-overlapping time windows (see \cref{fig:displacements}).

An alternative method for estimating the variance of the mean for time-correlated data is block averaging~\cite{Frenkel2023-ah}, where the input data is divided into non-overlapping sequential ``blocks'', and the set of averages calculated over each block are used for statistical analysis.
A popular form of block averaging is the block renormalisation method of Flyvbjerg and Peterson~\cite{FlyvbjergAndPetersen_JChemPhys1989}.

The Flyvbjerg--Peterson method starts with some input data, $A$.
If the elements of $A$ are uncorrelated, the variance of the mean can be estimated by rescaling the variance of $A$:
\begin{equation}
\varest{\xbar{A}}=\frac{\var{A}}{L_A-1},
\end{equation}
where $L_A$ is the number of elements in $A$.
If elements of $A$ are serially correlated, however, this estimator systematically underestimates the true variance of the mean of $A$, and, instead, providing only an approximate lower bound:
\begin{equation}
\varest{\xbar{A}}\geq\frac{\var{A}}{L_{A}-1}.
\end{equation}

\begin{figure}[tb]
  \centering
  \resizebox{9cm}{!}{\includegraphics*{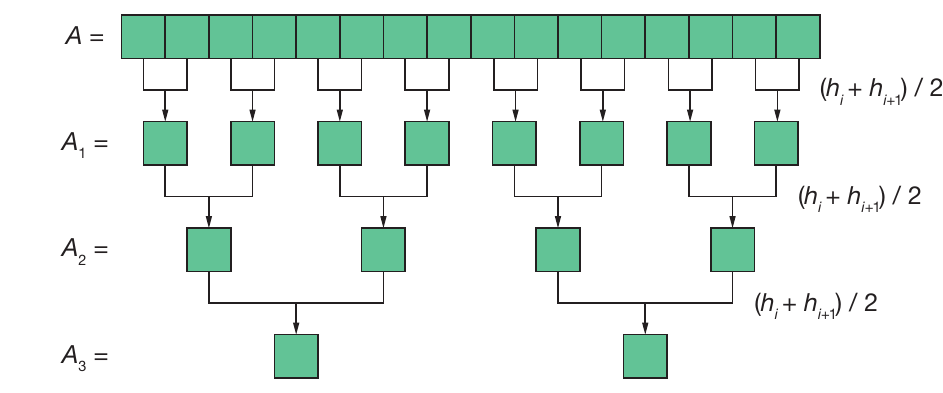}} 
    \caption{\label{fig:blocking}Schematic illustrating the application of repeated blocking operations with block-length 2 to generate a series of representations of the input data, $A$, $A_1$, $\ldots$, $A_n$, as used in the block renormalisation scheme of Flyvbjerg and Peterson~\cite{FlyvbjergAndPetersen_JChemPhys1989}.}
\end{figure}

The method proceeds by iteratively applying ``blocking'' operations.
The original dataset, $A$, is mapped to a new dataset $A_1$, by averaging over adjacent non-overlapping pairs of data in $A$ (see \cref{fig:blocking}).
The new dataset is half the length of the original: $L_{A_1} = \frac{1}{2}L_{A}$.
$A$ and $A_1$ have the same mean.
However, we can now rescale the variance of $A_1$ to obtain a tighter estimated lower bound for \var{\xbar{A}}:
\begin{equation}
\label{equ:blocking}
\var{\xbar{A}}\geq\frac{\var{A_1}} {L_{A_1}-1}.
\end{equation}
Under repeated blocking operations with a sufficiently large input dataset, $\var{A_n}/(L_{A_n}-1)$ tends to \var{\xbar{A}}.
In practice, sequential blocking steps are applied to the original dataset until $\var{A_n}/(L_{A_n}-1)$ reaches a plateau, and this plateau value is taken as the estimate for \var{\xbar{A}}.

The Flyvbjerg--Peterson block renormalisation method is particularly useful for estimating the variance of the mean in sequential correlated data where the correlation length is unknown; for example, when estimating thermodynamic averages from simulation trajectories.

\begin{figure*}[tb]
  \centering
  \resizebox{11cm}{!}{\includegraphics*{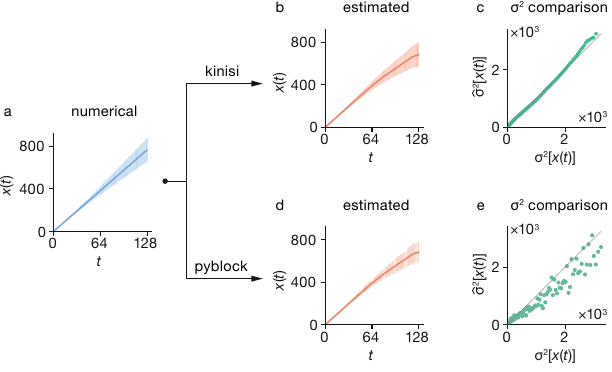}} 
    \caption{\label{fig:msd_blocking_comparison}Comparison of the numerical variance in observed MSD from multiple replica simulations (panel a), the estimated variance obtained by rescaling the variance in observed squared displacements from a single simulation (panels b and c) (Eqn.~7), and the estimated variance obtained from the block renormalisation method of Flyvbjerg and Peterson~\cite{FlyvbjergAndPetersen_JChemPhys1989}, as implemented in \textsc{pyblock}~\cite{spencer_pyblock_2020}. Panel (a) shows the mean observed MSD from \num{4096} simulations of \num{128} particles undergoing a 3D lattice random walk of 128 steps per particle, with error bars of $\pm2\var{\oMSDi}$. Panel (b) shows the MSD from just one simulation, with error bars of $\pm\varest{\oMSDi}$, obtained via Eqn.~7. Panel (d) shows the same one-simulation MSD, again with error bars of $\pm\varest{\oMSDi}$, obtained using the block renormalisation method of Flyvbjerg and Peterson~\cite{FlyvbjergAndPetersen_JChemPhys1989}. Panels (c) and (e) plot the numerical variance against the single-simulation estimated variances obtained with each method, as a function of timestep, $i$.}
    \script{msd_blocking.py}
\end{figure*}

Figure~\ref{fig:msd_blocking_comparison} shows a comparison of the variance rescaling method described in the main text with the Flyvbjerg--Petersen block renormalisation method when using both methods to estimate the $\var{\oMSDi}$ from simulation.
We applied both methods to a simulation of \num{128} particles undergoing a three-dimensional cubic-lattice random walk of \num{128} steps per particle. 

The variance rescaling method (Eqn.~7) shows close agreement between the estimated variance in the MSD from a single simulation and the true numerical variance obtained by sampling many equivalent simulations.
The estimated variance varies smoothly with timestep, $i$.
In the example shown, this method overestimates $\var{\oMSDi}$ at long times, which we attribute the the relatively small number of observed squared displacements that are used to estimate $\var{\oMSDi}$ in the large $i$ regime.

In comparison, the block renormalisation method also provides reasonable estimates of the true numerically determined variance, but is generally less accurate than the variance rescaling method.
This method exhibits more noise, with larger scatter in $\varest{\oMSDi}$ as the timestep $i$ changes.
This greater noise is not entirely surprising, since the block renormalisation method aims to independently estimate the correlation length of the input data numerically for each timestep $i$, while the variance rescaling method takes advantage of the known correlation length due to the way the MSD is computed (\cref{fig:displacements}).
The block renormalisation method also tends to underestimate $\var{\oMSDi}$, which can be attributed to its provision of an estimated lower bound for the variance of the mean of the input data (see \cref{equ:blocking}).

\begin{figure*}[tb]
  \centering
  \resizebox{10.0cm}{!}{\includegraphics*{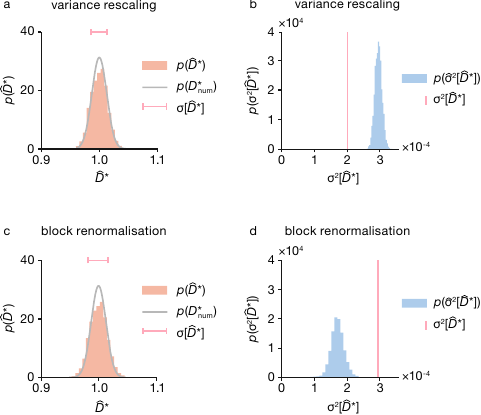}} 
    \caption{\label{fig:pyblock_comparison}(Panels (a) and (c)) Probability distributions of point-estimates $\prob{\D}$ obtained from \num{4096} individual random-walk simulations. Panel (a) shows data obtained using variance rescaling to estimated the variances, $\varest{\oMSDi}$, used to parametrise the model covariance matrix, $\mathbf{\Sigma^\prime}$. Panel (b) shows data obtained using block renormalisation to estimate the variances, $\varest{\oMSDi}$, used to parametrise the model covariance matrix.
    The grey lines show the distribution of point estimates, $\prob{\Dest}$, obtained using Bayesian regression with a mean vector and numerical covariance matrix derived from the complete dataset of all \num{4096} simulations. 
    The pink horizontal bar shows an interval of one standard deviation in $\prob{\Dest}$. 
    Panels (b) and (d) show the corresponding  probability distributions of estimated variances, $\varest{\Dest}$, 
    for individual random-walk simulations, compared to the true sample variances obtained using each method (pink vertical lines) 
    $\var{\Dest}$.
    }
    \script{pyblock.py}
\end{figure*}

In \cref{fig:pyblock_comparison} we compare the results of estimating $\D$ and $\var{\Dest}$ using our approximate Bayesian regression scheme, using a model covariance matrix, $\mathbf{\Sigma^\prime}$, as defined in Eqn.~6 in the main text, parameterised by $\varest{\oMSDi}$ computed by either variance rescaling or block renormalisation.

Both methods give unbiased estimates of $\D$ and similar distributions $\prob{\Dest}$.
However, the distribution $\prob{\Dest}$ obtained using variance rescaling is slightly narrower than that obtained using block renormalisation.
This difference stems from the greater stochastic noise in $\varest{\oMSDi}$ when using the block renormalisation method.
Consequently, the model covariance matrices, $\mathbf{\Sigma^\prime}$, parametrised from these estimates are themselves noisier and often difficult to condition, leading to numerical instabilities.
Both the increased noise and numerical instabilities when conditioning contribute to a wider distribution in $\prob{\Dest}$ when block renormalisation is used to estimate $\var{\oMSDi}$.

The choice of method used to estimate $\var{\oMSDi}$ also influences the estimate of the uncertainty in $\Dest$.
The variance rescaling method provides a good estimate for $\var{\Dest}$ but systematically overestimate the true uncertainty.
In contrast, the block renormalisation method provides a similarly good estimate but systematically \emph{underestimates} the true uncertainty in $\Dest$.
When estimating $\D$ from molecular dynamics simulations, we consider overestimation of $\var{\Dest}$ to always be preferable to underestimation.
Overestimated uncertainty in $\Dest$ can be addressed by collecting more data, for example, by performing longer simulations.
Conversely, underestimated uncertainty in $\Dest$ may lead to false confidence in the accuracy of $\D$ estimates, which can potentially lead to downstream errors in inferential reasoning or formal hypothesis testing.

\section{Bias in $\prob{\varest{\Dest}}$}
\label{sec:bias}
In the main manuscript, we present results for a set of \num{4096} 3D-lattice random walk simulations, each consisting of \num{128} particles undergoing \num{128} steps (Fig.~4).
Our approximate Bayesian regression scheme allows us to estimate the variance in $\Dest$, denoted as $\varest{\Dest}$, that would be obtained over a large number of repeat simulations.
This estimate is calculated from the variance of the marginal posterior distribution $\prob{\D|\model}$, which we derive from analysis of a single simulation trajectory.
As shown in Fig.~4d, our estimate for the population variance $\var{\Dest}$, obtained from a single simulation, aligns reasonably with the true value.
When considering the distribution of estimated variance, $\prob{\varest{\Dest}}$, however, we observe a systematic overestimation (bias) relative to the true value.

\begin{figure}[tb!]
    \centering
    \resizebox{\columnwidth}{!}{\includegraphics*{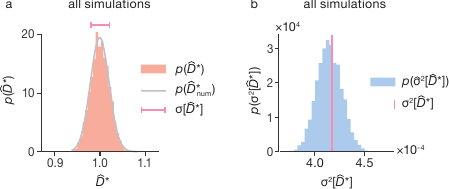}}
    \caption{
        (a) Probability distribution of point-estimates, $\prob{\Dest}$, obtained from \num{4096} individual random-walk simulations, using the numerical covariance matrix $\bm{\Sigma}_\mathrm{num}$.
        Each simulation has been analysed as in Fig.~4(a) and (b) to yield a single corresponding point estimate $\Dest$.
        The grey line shows the distribution of point estimates, $\prob{\Dest_\mathrm{num}}$, obtained using Bayesian regression with a mean vector and numerical covariance matrix derived from the complete dataset of all \num{4096} simulations.
        The pink horizontal bar shows an interval of one standard deviation in $\prob{\Dest}$.
        (b) Probability distribution of estimated variances, $\varest{\Dest}$, for individual random-walk simulations, using the numerical covariance matrix $\bm{\Sigma}_\mathrm{num}$, compared to the true sample variance (pink vertical line) $\var{\Dest}$.
        }
    \label{fig:true_cov}
    \script{true_cov.py}
\end{figure}

This bias arises from our use of estimated variances $\varest{\oMSDi}$ when parametrising the model covariance matrix $\bm{\Sigma^\prime}$.
Figure~\ref{fig:true_cov} presents equivalent results for $\prob{\Dest}$ and $\prob{\varest{\Dest}}$ for the same \num{4096} individual simulations, but calculated using a numerical covariance matrix, $\bm{\Sigma}_\mathrm{num}$ derived from all \num{4096} observed MSDs.
The resulting distribution $\prob{\varest{\Dest}}$ (see Fig.~\ref{fig:true_cov}b) is unbiased.
Furthermore, the distribution $\prob{\Dest}$ agrees even more closely with the numerically converged distribution obtained when combining data from all \num{4096} simulations (Fig~\ref{fig:true_cov}a), contrasting with the results presented in Fig.~4b, where our approximate Bayesian regression scheme yields a slightly broadened distribution due to the use of the long-time limit in the derivation of the analytical form for $\bm{\Sigma^\prime}$.

\section{Comparison of OLS, WLS, and GLS used to estimate $\D$ in \ce{Li7La3Zr2O12}}
\label{sec:llzo}
In the main manuscript, Fig.~1 shows example distributions of estimated self-diffusion coefficients, $\Dest$, calculated using OLS, WLS, and GLS estimators from MSD data from \num{4096} 3D lattice random walk simulations.
This figure shows that GLS gives a much narrower distribution of $\Dest$ than either OLS or WLS, and also allows the width of this distribution (characterised by $\var{\Dest}$) to be accurately estimated, in contrast with OLS and WLS, which both give estimates of $\var{\Dest}$ that significantly underestimate the true variance.

\begin{figure}[tb]
  \centering
  \resizebox{8cm}{!}{\includegraphics*{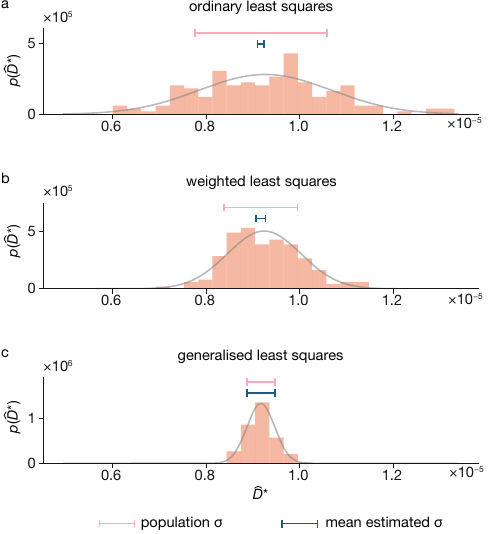}} 
    \caption{Example distributions of estimated self-diffusion coefficients, $\Dest$, calculated using (a) ordinary least squares (OLS), (b) weighted least squares (WLS), and (c) generalised least squares (GLS), from MSD data from \num{512} effective simulations of LLZO of \SI{\sim 25}{\pico\second} with \num{56} lithium ions. In each panel, the grey curve shows the best-fit normal distribution for the simulation data, the upper horizontal bar shows the standard deviation of this distribution, and the lower horizontal bar shows the average estimated standard distribution given by the analytical expression for $\var{\Dest}$ for each regression method.}
    \label{fig:glswlsols_llzo}
    \script{glswlsols_llzo.py}
\end{figure}

While the 3D lattice random walk represents an idealised model system, OLS and WLS show the same deficiencies when used to estimate $\D$ from MSD data from simulations of more complex ``real world'' systems.
\cref{fig:glswlsols_llzo} shows an equivalent comparison between OLS, WLS, and GLS applied to simulation data for the lithium solid electrolyte, \ce{Li7La3Zr2O12} (LLZO).
As for the idealised 3D lattice random walk data, OLS and WLS both give wider distributions of estimated diffusion coefficients, $\prob{\Dest}$, while also systematically underestimating the uncertainty in these estimates.
In contrast, GLS gives a narrower distribution of estimated values, and accurately estimates this uncertainty from single simulation data.

\end{document}